\documentclass[a4paper,twocolumn,english,twocolumns,prl]{revtex4}%
\usepackage{amsmath}
\usepackage{amsfonts}
\usepackage{amssymb}
\usepackage{graphicx}%
\providecommand{\U}[1]{\protect\rule{.1in}{.1in}}
\begin{document}
\title{Transport-induced correlations in weakly interacting systems}
\date{today}
\author{Guy Bunin$^1$, Yariv Kafri$^1$,  Vivien Lecomte$^2$, Daniel Podolsky$^1$ and Anatoli Polkovnikov$^3$}
\affiliation{$^1$ Department of Physics, Technion, Haifa 32000, Israel, $^2$ Laboratoire Probabilit\'es et Mod\`eles Al\'eatoires, UMR7599 CNRS, Universit\'e Pierre et Marie Curie \& Universit\'e Paris Diderot, 75013 \hbox{Paris,~France}, $^3$ Department of Physics, Boston University, Boston, MA 02215, USA.}

\begin{abstract}
We study spatial correlations in the transport of energy between two baths at different temperatures. To do this, we introduce a minimal model in which
energy flows from one bath to another through two subsystems. We show that the transport-induced
energy correlations between the two subsystems are of the same order as the
energy fluctuations within each subsystem. The correlations can be either
positive or negative and we give bounds on their values which are associated with
a dynamic energy scale. The different signs originate as a competition
between fluctuations generated near the baths, and fluctuations of the current
between the two subsystems. This interpretation sheds light on known results
for spatially-dependent heat and particle conduction models.
\end{abstract}
\maketitle


The physics of systems in and out of equilibrium can differ in dramatic ways. For example, in equilibrium, one-dimensional systems with short
range interactions cannot show long-range correlations at positive
temperatures. By contrast, in systems away from equilibrium
long-range correlations are known to form when there is a
steady-state current of a conserved quantity
\cite{schmitzcohen,dorfman_long_range,Derrida_review}. For diffusive systems
this has been calculated for models of particle and heat transport
\cite{SSEP_spohn,temp_long_range,GKR07,SchmittmannZia}, and measured in
heat-transport experiments \cite{dorfman_long_range,long_range_exp}.

%

\begin{figure}
[ptb]
\begin{center}
\includegraphics[
height=0.8912in,
width=1.7164in
]%
{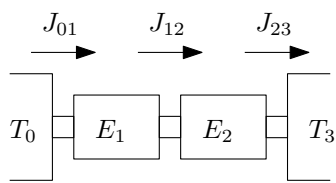}%
\caption{Heat conduction through a pair of systems. Subsystems 1 and 2 have
energies $E_{1}$ and $E_{2}$ respectively. The baths have temperatures
$T_{0},T_{3}$. The currents of energy between the different parts are
$J_{01},J_{12},J_{23}$, see figure.}%
\label{fig:3_systems_heat_cond}%
\end{center}
\end{figure}

In this paper, we study a minimal model for the formation of correlations
during the transport of a conserved quantity. We focus primarily on energy, but also show an example in which the same model is used to describe the transport of particles. The model
consists of two systems and two baths, arranged in a chain as shown in Fig.
\ref{fig:3_systems_heat_cond}. The total energy of the system is
$E_{\text{tot}}=E_{1}+E_{2}+E_{\text{int}}$, where $E_{1,2}$ denotes the
energy of the subsystems and $E_{\text{int}}$ is the interaction energy
between the two subsystems, and also between the subsystems and the baths. We
assume that the different components interact weakly, as is assured for
example when the interactions are short range and the links between the
components do not scale with system size. This means that the interactions
allow energy to flow, but can be neglected in any energetic calculation. In
equilibrium, \emph{i.e.}~when the temperatures of the two baths are equal,
$T_{0}=T_{3}$, the probability of subsystem 1 to be in state $s_{1}$ and
subsystem 2 to be in state $s_{2}$ is given to leading order in system size
by
\[
P\left(  s_{1},s_{2}\right)  =\frac{1}{Z}e^{-\beta E_{\text{tot}}}\approx
\frac{1}{Z}e^{-\beta E_{1}\left(  s_{1}\right)  }e^{-\beta E_{2}\left(
s_{2}\right)  }%
\]
where $\beta=1/T_{0}$ and $Z$ is the partition function. It follows
immediately that the energy correlations vanish%
\begin{equation}
\left\langle E_{1}E_{2}\right\rangle _{\beta}-\left\langle E_{1}\right\rangle
_{\beta}\,\left\langle E_{2}\right\rangle _{\beta}=0\ ,
\label{eq:equil_corr_is_zero}
\end{equation}
where $\langle\ldots\rangle_{\beta}$ denotes a thermal average.

In this Letter we use this minimal model to explain the mechanism by which long range correlations develop in systems away from equilibrium, \emph{i.e.} when $T_{0}\neq T_{3}$, despite the negligible interaction energies. These correlations are related to an energy scale associated with the
current transversing the system.  We apply our formalism to several examples, both classical and quantum, and show that the correlations can be both positive and negative.
We give a simple picture to explain the appearance and sign of
the correlation, and which also sheds light on continuous diffusive systems. 

The correlations discussed in this Letter can in principle be measured in
various systems. For example, correlations of the electric charge can be measured in double
quantum dot experiments \cite{DQD_review}, and in small analog electronic
circuits. The latter can also be used to measure heat fluctuations
\cite{ciliberto}. Such setups have been used in experimental studies of
fluctuation relations \cite{DQD,ciliberto}.

\emph{The model.--} The currents between the different components of the
system are denoted by $J_{01},J_{12},J_{23}$, see Fig.
\ref{fig:3_systems_heat_cond}. Conservation of energy implies that%
\begin{align}
\frac{dE_{1}}{dt} &  =J_{01}-J_{12}\ ,\nonumber\\
\frac{dE_{2}}{dt} &  =J_{12}-J_{23}\ .\label{eq:conserve}%
\end{align}
We consider simple dynamics, where the current fluctuations are modeled by
white noise. Separating fluctuations from average values, we write the
currents as
\begin{align}
J_{01} &  =A_{01}\left(  E_{1}\right)  +\sqrt{B_{01}\left(  E_{1}\right)
}\eta_{01}\nonumber\\
J_{12} &  =A_{12}\left(  E_{1},E_{2}\right)  +\sqrt{B_{12}\left(  E_{1}%
,E_{2}\right)  }\eta_{12}\nonumber\\
J_{23} &  =A_{23}\left(  E_{2}\right)  +\sqrt{B_{23}\left(  E_{2}\right)
}\eta_{23}\label{eq:J_langevin}%
\end{align}
where $A_{ij}\left(  E_{1},E_{2}\right)  =\overline{J_{ij}}\left(  E_{1}%
,E_{2}\right)  $, $\overline{\eta_{ij}}=0$, and $\overline{\eta_{ij}\left(
t\right)  \eta_{ij}\left(  t^{\prime}\right)  }=\delta\left(  t-t^{\prime
}\right)  $, for indices $ij=01,12,23$. Overbars denote averages at fixed
$E_{1},E_{2}$, and $B_{ij}$ are noise amplitudes.

The above model provides a good description of the dynamics when the energy 
flows are slow enough to allow subsystems 1 and 2 to constantly
relax to their microcanonical equilibria at energies $E_{1}$ and $E_{2}$. This
assures that the state of each subsystem is well-described by its energy. The
associated (microcanonical) inverse temperatures are $\beta_{i}\left(
E_{i}\right)  \equiv\partial S_{i}\left(  E_{i}\right)  /\partial E_{i}$ for
$i=1,2$, where $S_{i}\left(  E_{i}\right)  $ is the entropy of the $i$-th
subsystem. This time-scale separation also means that over the relaxation time
of the entire system to its non-equilibrium steady-state, the current performs
many independent fluctuations, and can therefore be modeled by white noise
about the average.

The physical information about the systems and the energy flow is contained in
the 6 functions $A_{ij}$ and $B_{ij}$ which as we show below can be obtained for 
specific models. Under some conditions the function $A_{ij}$ and $B_{ij}$ are related. Specifically,  let
 $w$ be the energy flow within the memory time of the noise (of the
order of the subsystems' relaxation time, or shorter). If $\left(
\Delta\beta\right)  ^{2}\overline{w^{3}}_{c}\ll\overline{w}$, where
$\overline{w^{3}}_{c}$ denotes the third cumulant and $\Delta\beta$ is the
largest of $\left\vert \beta_{1}-\beta_{0}\right\vert ,\left\vert \beta
_{2}-\beta_{1}\right\vert $ and $\left\vert \beta_{3}-\beta_{2}\right\vert $, then the following generalized
fluctuation-dissipation relation holds%
\begin{equation}
2A_{ij}=\left(  \beta_{j}-\beta_{i}\right)  B_{ij}\ . \label{eq:FD}%
\end{equation}
This follows by taking a cumulant expansion of the exchange fluctuation
relation \cite{fluct_heat_exchange}, and neglecting cumulants of order 3 and
higher, according to condition 2 above, as discussed in detail in
Ref.~\cite{our_coupled_sys} (for a different approach, see Ref.~\cite{Tikhonenkov}). It holds in general when the temperature differences are small
(namely in the standard linear response regime) but can hold more generally in some cases
(see Ref.~\cite{our_coupled_sys} for examples). In what follows we use this relation when applicable.

Finally, note that Eqs.~(\ref{eq:conserve}%
-\ref{eq:J_langevin}) describe~\cite{ILW09}, for a specific form of functions
$A_{ij}$ and $B_{ij}$, the dynamics of local energy transfers in a stochastic
form~\cite{BGL05,GKR07} of the discrete Kipnis, Marchioro, and Presutti
model~\cite{KMP}.  We relate them to other models below.

\emph{Steady-state fluctuations.--} With the model defined, we turn to study
the energy fluctuations at its steady-state. At the steady-state,
$\left\langle dE_{1}/dt\right\rangle =\left\langle dE_{2}/dt\right\rangle =0$,
where angular brackets denote averages over the steady-state probability
distribution. From Eqs. (\ref{eq:conserve}) and (\ref{eq:J_langevin}), this
implies $\left\langle A_{01}\right\rangle =\left\langle A_{12}\right\rangle
=\,\left\langle A_{23}\right\rangle $.

As in equilibrium, fluctuations are expected to scale as $N^{1/2}$, where $N$
is the system size. This will be shown self-consistently below. This motivates
an expansion in small energy fluctuations $\delta E_{1}\equiv E_{1}%
-\,\left\langle E_{1}\right\rangle ,\delta E_{2}\equiv E_{2}-\,\left\langle
E_{2}\right\rangle $.

Using this we first find $\left\langle E_{i}\right\rangle $. To lowest order
in $\delta E_{i}$, we have $\left\langle A_{ij}\left(  E_{1},E_{2}\right)
\right\rangle =A_{ij}\left(  \left\langle E_{1}\right\rangle ,\left\langle
E_{2}\right\rangle \right)  .$ Therefore at $\delta E_{1}=\delta E_{2}=0$,%
\begin{align}
A_{01}\left(  \left\langle E_{1}\right\rangle \right)  -A_{12}\left(
\left\langle E_{1}\right\rangle ,\left\langle E_{2}\right\rangle \right)   &
=0\ ,\nonumber\\
A_{12}\left(  \left\langle E_{1}\right\rangle ,\left\langle E_{2}\right\rangle
\right)  -A_{23}\left(  \left\langle E_{2}\right\rangle \right)   &
=0\ ,\label{eq:Asolve}%
\end{align}
which can be solved to obtain $\left\langle E_{1}\right\rangle $ and
$\left\langle E_{2}\right\rangle $.

We next expand Eq.~(\ref{eq:conserve}) to leading order in the fluctuations to
obtain:
\begin{equation}
\frac{d}{dt}\binom{\delta E_{1}}{\delta E_{2}}=\mathbf{R}\binom{\delta E_{1}%
}{\delta E_{2}}+\mathbf{\Sigma}\left(
\begin{array}
[c]{c}%
\eta_{01}\\
\eta_{12}\\
\eta_{23}%
\end{array}
\right)  \ . \label{eq:linear_langevin}%
\end{equation}
Here $\mathbf{{R}}$ is a $2\times2$ matrix with elements $R_{ij}=\left.
\partial_{E_{j}} X_{i}\right|  _{E_{1,2}=\langle E_{1,2} \rangle},$ where
$X_{1}=A_{01}-A_{12}$ and $X_{2}=A_{12}-A_{23}$. The matrix $\mathbf{R}$
controls the relaxation of fluctuations, as $d\left(  \overline{\delta E_{i}%
}\right)  /dt=\sum_{j}R_{ij}\overline{\delta E_{j}}$. ${\mathbf{\Sigma}}$ is a
$3\times2$ matrix which specifies the strength of the fluctuations. Since the
numbers $B_{ij}$ do not generally vanish at $E_{1}=\left\langle E_{1}%
\right\rangle ,E_{2}=\left\langle E_{2}\right\rangle $ we have to lowest order
in the fluctuations
\[
\mathbf{\Sigma}\equiv\left(
\begin{array}
[c]{ccc}%
\sqrt{B_{01}^{s}} & -\sqrt{B_{12}^{s}} & 0\\
0 & \sqrt{B_{12}^{s}} & -\sqrt{B_{23}^{s}}%
\end{array}
\right)  \ ,
\]
with $B_{ij}^{s}=B_{ij}\left(  \left\langle E_{1}\right\rangle ,\left\langle
E_{2}\right\rangle \right)  $.

Finally, the energy correlation matrix $C_{ij}=\langle E_{i} E_{j}\rangle-
\langle E_{i} \rangle\langle E_{j} \rangle$ can be obtained by solving the
Lyapunov equation \cite{gardiner}%
\begin{equation}
\mathbf{RC+CR}^{T}=-\mathbf{Q}\ , \label{eq:Lyapunov}%
\end{equation}
where $\mathbf{Q\equiv\Sigma\Sigma}^{T}$. In general, away from equilibrium
the off-diagonal $C_{12}$ does not vanish. In what follows, we will
demonstrate this explicitly in two simple models.

It follows from Eq. (\ref{eq:FD}) that$\ A_{ij}$ and $B_{ij}$ have the same
scaling with system size. $\mathbf{R}$ is composed of derivatives of the type
$\partial A_{ij}/\partial E_{k}$, and therefore scales as $N^{-1}$, where $N$
is the system size. From Eq. (\ref{eq:Lyapunov}) it follows that
$\mathbf{C}\sim N$, as in equilibrium. The fluctuations in the energy then
scale as $N^{1/2}$ and justify self-consistently the expansion in small
$\delta E_{1},\delta E_{2}$.

\emph{Linear $A$ model.--} Consider a simple model, where subsystems 1 and 2
are identical, with microcanonical temperatures $T(E)$. $E\left(  T\right)  $
will denote the inverse of this function. In addition, for the dynamics we
take%
\begin{align}
A_{01} &  =\gamma\left[  \varepsilon\left(  T_{0}\right)  -\varepsilon
_{1}\right]  \ ,\nonumber\\
A_{12} &  =\gamma\left[  \varepsilon_{1}-\varepsilon_{2}\right]
\ ,\nonumber\\
A_{23} &  =\gamma\left[  \varepsilon_{2}-\varepsilon\left(  T_{3}\right)
\right]  \ .\label{eq:A_lin_form}%
\end{align}
Here $\varepsilon$ are the energy densities in the baths ($\varepsilon(T_{0})$
and $\varepsilon(T_{3})$) and in the subsystems ($\varepsilon_{1,2}=E_{1,2}%
/N$); and $\gamma$ is a rate constant, setting a time scale in the model. From
the definition of $\mathbf{R}$%
\[
\mathbf{R}=\frac{\gamma}{N}\left(
\begin{array}
[c]{cc}%
-2 & 1\\
1 & -2
\end{array}
\right)  \ .
\]
The solution to Eq. (\ref{eq:Lyapunov}) then reads
\begin{equation}
\mathbf{C}=\frac{N}{24\gamma}\left(
\begin{array}
[c]{cc}%
7B_{01}+4B_{12}+B_{23} & 2B_{01}-4B_{12}+2B_{23}\\
2B_{01}-4B_{12}+2B_{23} & B_{01}+4B_{12}+7B_{23}%
\end{array}
\right)  .\label{eq:Lyap_solution}%
\end{equation}
Note that at equilibrium $B_{01}=B_{12}=B_{23}$, so that ${\mathbf{C}}$ is
diagonal,
as expected from Eq. (\ref{eq:equil_corr_is_zero}).

Interestingly, we see that the cross-correlation $C_{12}$ can be both positive
or negative, depending on the sign of $B_{01}-2B_{12}+B_{23}$. We suggest the
following simple interpretation: current fluctuations transfer energy along
the different bonds. The fluctuations relax by the average dynamics $d\left(
\overline{\delta E_{i}}\right)  /dt=\sum_{j}R_{ij}\overline{\delta E_{j}}$.
Consider a current fluctuation near a bath, say in $J_{01}$, which changes the
energies by $\left(  \delta E_{1},\delta E_{2}\right)  \propto\left(
1,0\right)  $. This fluctuation relaxes according to the sum of the modes of
$\mathbf{R}$, $\left(  \delta E_{1},\delta E_{2}\right)  \propto e^{-\gamma
t/N}\left(  1,1\right)  -e^{-3\gamma t/N}\left(  -1,1\right)  $, for which
$\delta E_{1}$ and $\delta E_{2}$ have the same sign. On the other hand, a
fluctuation in the central bond $J_{12}$ changes energies by $\left(  \delta
E_{1},\delta E_{2}\right)  \propto\left(  -1,1\right)  $, which decays as
$\left(  \delta E_{1},\delta E_{2}\right)  \propto-e^{-3\gamma t/N}\left(
-1,1\right)  $. Therefore, current fluctuations near the baths promote
positive correlations between the two subsystems, while current fluctuations
in the center of the system contribute to negative correlation between the
subsystems. The expression $C_{12}\propto B_{01}-2B_{12}+B_{23}$ reflects the
positive effect of the boundary noise $B_{01},B_{23}$ and the negative effect
of the bulk noise $B_{12}$.

To demonstrate  the different possible behaviors in different models consider first a specific example of the linear $A$ model in which the
subsystems are ideal gases with $E=C_{v}T$, with $C_{v}$ a constant specific
heat and that equation Eq. (\ref{eq:FD}) holds.  This model of ideal gases satisfying Fourier's law is perhaps the simplest phenomenological model for energy transfer.  To obtain $B_{01}^{s},B_{12}^{s},B_{23}^{s}$ we solve Eq.
(\ref{eq:Asolve}) and use Eq. (\ref{eq:FD}). Substituting into Eq.
(\ref{eq:Lyap_solution}) we find%
\[
C_{12}=\frac{C_{v}}{27}(T_{0}-T_{3})^{2}%
\]
Thus $C_{12}$ is positive for any $T_{0}\neq T_{3}$, and proportional to the
system size. 

\begin{figure}
[ptb]
\begin{center}
\includegraphics[
height=1.8137in,
width=2.6749in
]%
{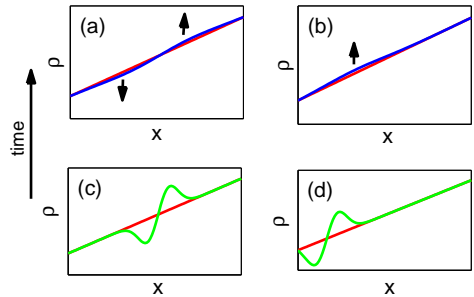}%
\caption{In continuum models with constant diffusivity, a current fluctuation
in the center of the system (c) relaxes to a density fluctuation (a)\ which
gives a negative contribution to spatial correlations. In contrast, a current
fluctuation near the boundaries (d), adds a positive contribution to spatial
correlations (b).}%
\label{fig:corr_decay}%
\end{center}
\end{figure}

This model can be considered as a \textquotedblleft boxed\textquotedblright%
\ version of standard models for heat  or particle conduction with spatial dependence,
known to have positive correlations in the continuum limit
\cite{temp_long_range}. Indeed, a similar line of argumentation applies to
continuum diffusive systems, which conduct particles or heat. If a fluctuation
in the energy density $E\left(  x,t\right)  $ decays according to a simple
diffusion $\partial_{t}E=D\nabla^{2}E$, with $D$ a constant diffusion coefficient,  a current fluctuation near the baths
will contribute to positive correlations, while current fluctuations at the
bulk will contribute to negative correlations. Therefore models with stronger
noise near the boundary will have positive correlations, while models with
stronger bulk noise will have negative correlations, as illustrated in Fig.~\ref{fig:corr_decay}. This gives a simple
picture of the positive correlations found in standard heat conduction
models \cite{temp_long_range,GKR07} as opposed to negative correlation found, for
example, in the simple symmetric exclusion process (SSEP) \cite{SSEP_spohn,DEHP93}, a
standard particle conduction model.

Indeed, a boxed version of the SSEP can be constructed as follows.  Consider a boxed model in which $\varepsilon_i$ describes the density of particles in box $i$.  A particle then hops from subsystem $i$ to $j$ with rate $\gamma \varepsilon_i(1-\varepsilon_j)$ where $1$ is the maximal density of particles that each subsystem can contain.  This leads \cite{gardiner} 
to a linear $A=\gamma(\varepsilon_i-\varepsilon_j)$ model with a variance in the particle transfer between boxes 
$B_{ij}=\gamma \left(\varepsilon_i(1-\varepsilon_j)+\varepsilon_j(1-\varepsilon_i)\right).$
Using the procedure outlined above we find here
\begin{equation}
C_{12}=\frac{- N}{27}(\varepsilon_0-\varepsilon_3)^2\;.
\end{equation}
Note that now the correlations are negative as opposed to positive in the previous example.


We next turn to calculate the correlations in two quantum examples. In the first we consider coupled Fermi gases.  In the second, treated in the Supplementary Material, we study systems which exchange energy through blackbody radiation. In both
cases the correlations turn out to be positive.

\emph{Ideal Fermi gas.--} We now derive the correlation function starting from
a quantum mechanical model of weakly interacting Fermi gases. To derive the
average energy flow, we first consider two Fermi gases in adjacent boxes, with
Hamiltonian $H=H_{c}+H_{d}+H^{\prime}$, where $H_{c} =\sum_{k}\epsilon_{k}^{c}c_{k}^{\dagger}c_{k}$,  $H_{d}  =\sum_{k}\epsilon_{k}^{d}d_{k}^{\dagger}d_{k}$, and $H^{\prime} =\lambda\,c_{0}^{\dagger}c_{0}d_{0}^{\dagger}d_{0}.$
In box c (d) we label fermion operators by $c,c^{\dagger}$ ($d,d^{\dagger}$),
with dispersion $\epsilon_{k}^{c}$ ($\epsilon_{k}^{d}$). In addition, there is
a density-density interaction between the $c$ and $d$ fermions, localized at
the contact point (\textquotedblleft0\textquotedblright) between the two
boxes. This interaction allows for the exchange of energy between the boxes,
but does not permit particle exchange between them.

For small $\lambda$ we can treat $H^{\prime}$ as a weak perturbation. Its
effect is to scatter a pair of electrons $\{c_{k_{2}},d_{k_{4}}\}$ to a new
pair $\{c_{k_{1}},d_{k_{3}}\}$. Each such process leads to a transfer of
energy $\epsilon_{k_{3}}-\epsilon_{k_{4}}$.
The rate for this process is computed using Fermi's golden rule, leading
to an average energy current $A_{cd}=f^{(1)}_{cd}$ and fluctuations $B_{cd}=f^{(2)}_{cd}$, where
\begin{align}
f^{(m)}_{cd}=& \frac{2\pi|\lambda|^{2}}{\hbar} \sum_{k_{1}\ldots k_{4}}
(\epsilon_{k_{3}}-\epsilon_{k_{4}})^{m}\,
n^{c}_{k_{2}}n^{d}_{k_{4}}\\
&\,\times(1-n^{c}_{k_{1}})(1-n^{d}_{k_{3}})\delta\left(  \epsilon_{k_{2}%
}+\epsilon_{k_{4}}-\epsilon_{k_{1}}-\epsilon_{k_{3}}\right)\nonumber
\end{align}
where $n^{c}_{k}=\frac{1}{\exp\frac{\epsilon_{k}-\mu_{c}}{T_{c}}+1}$ and
$n^{d}_{k}=\frac{1}{\exp\frac{\epsilon_{k}-\mu_{d}}{T_{d}}+1}$ are Fermi
occupation functions and $\mu_{c,d}$ are chemical potentials. Note that we allow for different densities and
temperatures in the two boxes.

We can replace the sums over $k$ by integrals over energy, $\sum_{k}\to\int
d\epsilon\, \nu(\epsilon)$. Furthermore, provided that $T_{c},T_{d}\ll T_{F}$,
the Fermi temperature, we can ignore the energy dependence of the density of
states, and replace $\nu(\epsilon)\to\nu_{c,d}$ for each one of the boxes. We
then obtain
\begin{align}
A_{cd}=\kappa_{cd}(T_{c}^{4}-T_{d}^{4}) \label{eq:fermi}%
\end{align}
where $\kappa_{cd}= \frac{2\pi^{5}}{15\hbar}|\lambda|^{2}\nu_{c}^{2}\nu
_{d}^{2}.$ $B_{cd}$ is found numerically to be well-described by
\begin{eqnarray}
B_{cd}\approx 4\kappa_{cd}(T_c^5+T_d^5)
\label{eq:Bcd}
\end{eqnarray}
which satisfies Eq.~(\ref{eq:FD}) when $T_c\approx T_d$.
In what follows we will consider the case where the subsystems
and the baths are all described by Fermi gases, and we will take $\kappa
_{cd}=\kappa$ to be constant across all junctions.

Using Eqs.~(\ref{eq:fermi}) and (\ref{eq:Asolve}) we solve for the average
energies to find
$T_{1}  =((2T_{0}^{4}+T_{3}^{4})/3)^{\frac{1}{4}}$ and
$T_{2}  =((T_{0}^{4}+2T_{3}^{4})/3)^{\frac{1}{4}}$.
From this, following the procedure described above, and using the expression for $E$ to lowest order in $T/T_{F}$,
$E_{i}=\frac{3}{5}NT_{F}\left(  1+\frac{5\pi^{2}}{12}\left(  T_{i}/T_{F}\right)  ^{2}\right),$
we find
\begin{align}
C_{12}  &  =\frac{\pi^{2}N\left(B_{01}-2B_{12}+B_{23} \right)}{48\kappa T_F  (T_{1}^{2}+T_{2}^{2})} 
\end{align}
whose sign depends on the convexity of $B_{cd}$.  Using Eq.~(\ref{eq:Bcd}) we find that correlations are always positive and increase with the temperature difference between the two baths.

\emph{Bound on the energy correlations.--} It is natural to ask how the
magnitude of the correlations is related to the different energy scales in the
problem: $k_{B}T_{i}$ for $i=0..3$, and a dynamic energy scale $\left\langle
J\right\rangle \tau$, composed of the average current $\left\langle
J\right\rangle $ multiplied by the relaxation time of fluctuation to the
steady-state, $\tau$. The last energy scale is directly related to the
non-equilibrium steady state.

In the supplementary materials we consider systems in which Eq.~(\ref{eq:FD}) holds, and show that, if the average currents grow with
the difference between $E_{i}$ and $E_{j}$ (i.e. $\partial_{E_{i}}A_{ij}>0$
and $\partial_{E_{j}}A_{ij}<0$), then
\begin{equation}
\left\vert C_{12}\right\vert \leq18\,\left\langle J\right\rangle \tau
\max_{i=0,1,2}\left(  \left\vert \beta_{i}-\beta_{i+1}\right\vert
^{-1}\right)  \ .\label{eq:C12_bound}%
\end{equation}
This gives an upper bound on the size of the correlations. The bound is
proportional to the dynamic energy scale $\langle J\rangle\tau$, demonstrating
how correlations disappear upon approach to equilibrium. 

The bound above can be understood as follows. Correlations have units of
energy squared, and vanish at equilibrium. One therefore expects them to scale
as $\left\langle J\right\rangle \tau$, multiplied by an energy scale derived
from the energies $k_{B}T_{i}$. This is expressed by the Lyapunov equation Eq.
(\ref{eq:Lyapunov}). The matrix $\mathbf{R}$ sets the time scale $\tau$ for
relaxation, with $\tau^{-1}=\lambda_{\max}$, the larger of the eigenvalues of
$\mathbf{R}$. The elements of the matrix $\mathbf{Q}$ are sums of $B_{ij}$
terms. According to Eq. (\ref{eq:FD}), $B_{ij}=2A_{ij}/\left(  \beta_{i}%
-\beta_{j}\right)  \,$, these are proportional to the averaged currents
$A_{ij}$ multiplied by a combination of the energy scales $k_{B}T_{i}$. The
solution to the Lyapunov equation therefore scales as $\left\langle
J\right\rangle \tau$ multiplied by a combination of the energies $k_{B}T_{i}$.


Interestingly, the energy scale $\left\langle J\right\rangle \tau$ also
appears in an exact relation for the non-equilibrium fluctuations in a simpler
model, where a single system is connected to two heat baths.
In Ref.~\cite{our_coupled_sys} it was shown that when Eq.~(\ref{eq:FD}) holds, energy fluctuations in the system are given by $\left\langle
J\right\rangle \tau\frac{\beta_{2}-\beta_{1}}{\left(  \beta_{0}-\beta
_{1}\right)  \left(  \beta_{2}-\beta_{0}\right)  }\ .
$
%


Finally, we comment that the large deviation functional (LDF) of the energy or density profile in
non-equilibrium systems has recently been the subject of close attention
\cite{Derrida_review,BGL05,ILW09,BertiniPRL,TKL_long,our_numerics,ASEP_transition,ours_short,lr-corr_ph-trans}.
In particular it has been shown that the non-locality of the LDF is directly related to the non-equilibrium long-range correlations~\cite{Derrida_review,BSGJL07}.
The model studied here, being low dimensional, might serve as a good
template for understanding their general properties, in particular, in light
of the insights gained on the sign of the correlations in driven diffusive systems.
To date, the LDF has been determined for a few models in the macroscopic limit~\cite{BGL05,ED04} and obeys a variational principle which remains to be understood for generic systems.

\emph{Acknowledgements.--} We are very grateful to Luca D'Alessio for many useful discussions and important comments.  We would like to thank support of BSF and ISF grants, the 
EU under grant agreement no.~276923,--- MC--MOTIPROX, NSF DMR-0907039 and the SCHePS Paris~7 interdisciplinary project.

%

\appendix

\section{Appendix A: Coupled blackbodies}

Let's consider two coupled cavities at different temperatures.  Each cavity has a thermal gas of photons, and the photons can leak from one cavity to the other at the interface between the two.  The Hamiltonian is
\begin{eqnarray}
H=\sum_{\mathbf{k},\alpha}  \hbar c k a_{\mathbf{k},\alpha}^\dagger a_{\mathbf{k},\alpha}+\sum_{\mathbf{k},\alpha} \hbar c k b_{\mathbf{k},\alpha}^\dagger b_{\mathbf{k},\alpha}+H_{\rm int}.
\end{eqnarray}
Here $a,a^\dagger$ and  $b,b^\dagger$ denote creation and annihilation operators of photons with momentum ${\mathbf{k}}$ and polarization $\alpha$ in each of the two cavities and $k=|{\mathbf{k}}|$.  We impose reflecting boundary conditions at the surface $z=0$ separating the two cavities, so that $k_z$ in the sums is restricted to positive values.  The interaction term
\begin{eqnarray}
H_{\rm int}=\lambda \hbar c \sum_{\alpha=1}^2\int dx\,dy\, a^\dagger_{(x,y,0),\alpha}b^{\,}_{(x,y,0),\alpha}+h.c.
\label{eq:hint}
\end{eqnarray}
effectively converts outgoing $a$ photons from the first cavity into $b$ photons in the second cavity. 
The prefactor $\lambda$ plays the role of the transmission through the barrier separating the two systems and we assume that the barrier does not radiate, i.e. it is effectively at zero temperature. Usual blackbody radiation corresponds to $\lambda=1$. 

To calculate the energy transfer between the two systems we will use Fermi Golden rule treating $H_{int}$ as a perturbation.  Then, the matrix element squared for a photon of a given polarization with momentum ${\mathbf{k}}=(k_x,k_y,k_z)$ in box $a$ to transfer to box $b$ with momentum ${\mathbf{k}}'=(k_x,k_y,k_z')$ is
\begin{eqnarray}
\frac{|\lambda|^2\hbar^2c^2}{L^2} n^{(a)}_{\mathbf{k}} (1+n^{(b)}_{{\mathbf{k}}'})
\end{eqnarray}
where $n_{\mathbf{k}}^{(a,b)}=1/(e^{-\beta_{a,b} \hbar c k}-1)$ are the Bose occupation factors for the two cavities, and $L$ is the linear size of the cavities along the contact surface.  Each such process leads to an energy transfer $\hbar c k$.  Then, using the Fermi golden rule and summing over initial and final states, we find the energy transfer rate from box $a$ to $b$ to be
\begin{multline}
W_{a\to b}=\frac{|\lambda|^2\hbar c^2 L^2}{(2\pi)^3}\sum_{\alpha=1}^2\int d^3 {\mathbf{k}} \int dk_z'  \\
\hbar c k\,n^{(a)}_{\mathbf{k}} (1+n^{(b)}_{{\mathbf{k}'}})\delta(\hbar c k-\hbar c k').\nonumber
\end{multline}
Taking into account that $k_z> 0$ and $k_z'>0$, and using the fact that $\delta(\hbar c k-\hbar c k')=\frac{k_z}{\hbar c k}\delta(k_z-k_z')$, we obtain 
\begin{eqnarray}
W_{a\to b}=\frac{|\lambda|^2\hbar c^2 L^2}{4\pi^2}\int_0^\infty dk\,k^3  n^{(a)}_k (1+n^{(b)}_{k}).
\end{eqnarray}
This reduces to the usual result for blackbody radiation when $\beta_b=\infty$ and $\lambda=1$.

The net energy transfer rate from $a$ to $b$ is then
\begin{eqnarray}
A_{ab}=W_{a\to b}-W_{b\to a}=|\lambda|^2\sigma (T_a^4-T_b^4) 
\end{eqnarray}
where $\sigma=\frac{\pi^2 L^2}{60c^2\hbar^3}$ is the Stefan constant.  Note that since the energy density of a blackbody is proportional to $T^4$ the black body radiation results in a linear $A$ model. We can compute the energy fluctuations in a similar fashion:
\begin{multline}
B_{ab}= \frac{|\lambda|^2\hbar c^2 L^2}{(2\pi)^3} \sum_{\alpha=1}^2 \int d^3 {\mathbf{k}} \int dk_z' \, (\hbar c k)^2\,\delta(\hbar c k-\hbar c k') \nonumber\\
 \times \left[n^{(a)}_{\mathbf{k}} (1+n^{(b)}_{{\mathbf{k}}'})+n^{(b)}_{\mathbf{k}} (1+n^{(a)}_{{\mathbf{k}'}})\right] \nonumber \\
= \frac{|\lambda|^2\hbar^2 c^3 L^2}{(2\pi)^2}\int dk\, k^4   \left[n^{(a)}_k (1+n^{(b)}_k)+n^{(b)}_k (1+n^{(a)}_k)\right] \nonumber
\end{multline}
Numerically we find that this expression is well approximated by
\begin{eqnarray}
B_{ab}=8|\lambda|^2 \sigma \left[\left(T_aT_b\right)^{5/2}+\frac{45}{\pi^4}\left(T_a^{5/2}-T_b^{5/2}\right)^2\right]
\end{eqnarray}
which satisfies Eq.~(\ref{eq:FD}) when $T_a\approx T_b$.  Applying our formalism in a similar fashion to that described in the discussion of the Fermi gas, we find that the correlations are always positive.

As a side remark we note that the same result for the black body radiation can be obtained using a different type of perturbation
\begin{equation}
H_{int}= \frac{\lambda \hbar c}{i}\sum_{\alpha=1}^2\int dx\,dy\, \left. a^\dagger_{(x,y,0),\alpha} \partial_z b^{\,}_{(x,y,z),\alpha}\right|_{z=0}+h.c.
\end{equation}
At $\lambda=1$ this perturbation is nothing but the energy flux operator of photons. The easiest way to check that this perturbation gives the same result as Eq.~(\ref{eq:hint}) is to discretize the Hamiltonian along the $z$-direction
\begin{equation}
H_{int}=\lambda {\hbar c\over i d} \sum_{\alpha=1}^2\int dx\,dy\,  \left[a^\dagger_{(x,y,0),\alpha}  b^{\,}_{(x,y,d),\alpha}-h.c.\right],
\end{equation}
where $d$ is the lattice spacing and and then make the gauge transformation $b\to be^{-i\pi/2}$. This gauge transformation obviously does not affect $H_0$ while $H_{int}$ reduces to Eq.~(\ref{eq:hint}) in the continuum limit, using $1/ d \to \delta(z)$.  Hence we recover the equivalence of the two choices for the perturbation.

\section{Appendix B: Derivation of Eq. (\ref{eq:C12_bound})}

Here we derive the bound given in Eq. (\ref{eq:C12_bound}). We assume that
$\partial_{E_{i}}A_{ij}>0$ and $\partial_{E_{j}}A_{ij}<0$, meaning that the
average currents grow when the difference between $E_{i}$ and $E_{j}$ grows.
The matrix $\mathbf{R}$ is then given by%
\[
\mathbf{R}=\left(
\begin{array}
[c]{cc}%
-\gamma_{1}-\gamma_{2} & \gamma_{3}\\
\gamma_{2} & -\gamma_{3}-\gamma_{4}%
\end{array}
\right)  \ ,
\]
where $\gamma_{1}=-\partial_{E_{1}}A_{01}$, $\gamma_{2}=\partial_{E_{1}}%
A_{12}$, $\gamma_{3}=-\partial_{E_{2}}A_{12}$, and $\gamma_{4}=\partial
_{E_{2}}A_{23}$ are all positive numbers. As $\left\vert \operatorname{Tr}%
\left(  \mathbf{R}\right)  \right\vert =\left\vert \gamma_{1}+\gamma
_{2}+\gamma_{3}+\gamma_{4}\right\vert $ then $\max\gamma\equiv\max\left\{
\gamma_{i}\right\}  \leq\left\vert \operatorname{Tr}\left(  \mathbf{R}\right)
\right\vert $\textbf{.}

The correlation is given by
\begin{align*}
\text{\thinspace}C_{12}  &  =\frac{B_{23}(\gamma_{1}+\gamma_{2})\gamma
_{3}+B_{01}\gamma_{2}(\gamma_{3}+\gamma_{4})}{2\operatorname{Tr}\left(
\mathbf{R}\right)  \det\left(  \mathbf{R}\right)  }\\
&  -\frac{B_{12}\left(  \gamma_{2}\gamma_{4}+\gamma_{1}\gamma_{3}+2\gamma
_{1}\gamma_{4}\right)  }{2\operatorname{Tr}\left(  \mathbf{R}\right)
\det\left(  \mathbf{R}\right)  }%
\end{align*}
so that
\begin{align*}
\left\vert C_{12}\right\vert  &  \leq\frac{9\max\left(  B_{ij}\right)  \left(
\max\gamma\right)  ^{2}}{2\left\vert \operatorname{Tr}\left(  \mathbf{R}%
\right)  \right\vert \det\left(  \mathbf{R}\right)  }\\
&  \leq9\,\left\langle J\right\rangle \frac{\left\vert \operatorname{Tr}%
\left(  \mathbf{R}\right)  \right\vert }{\det\left(  \mathbf{R}\right)  }%
\max_{i=0,1,2}\left(  \left\vert \beta_{i}-\beta_{i+1}\right\vert
^{-1}\right)
\end{align*}
where in the second equality we used Eq. (\ref{eq:FD}), that at the steady
state $\left\langle J\right\rangle =A_{01}=A_{12}=A_{23}$, and that $\left(
\max\gamma\right)  ^{2}\leq\left[  \operatorname{Tr}\left(  \mathbf{R}\right)
\right]  ^{2}$. As $\left\vert Tr\left(  \mathbf{R}\right)  \right\vert
/\det\left(  \mathbf{R}\right)  =\lambda_{1}^{-1}+\lambda_{2}^{-1}\leq2\tau$,
one obtains Eq. (\ref{eq:C12_bound})%
\[
\left\vert C_{12}\right\vert \leq18\,\left\langle J\right\rangle \tau
\max_{i=0,1,2}\left(  \left\vert \beta_{i}-\beta_{i+1}\right\vert
^{-1}\right)  \ .
\]

\bigskip


\begin{thebibliography}{99}                                                                                               %

\bibitem {Derrida_review}B. Derrida, J. Stat. Mech. P07023 (2007)

\bibitem {schmitzcohen}R. Schmitz and E.D.G. Cohen, J. Stat. Phys. \textbf{39},
285 (1985)

\bibitem {dorfman_long_range}J. R. Dorfman, T. R. Kirkpatrick and J. V.
Sengers, Annu. Rev. Phys. Chem. \textbf{45}, 213-239 (1994)

\bibitem {SSEP_spohn}H. Spohn, J. Phys. A: Math. Gen. \textbf{16}, 4275 (1983)

\bibitem {temp_long_range}A. L. Garcia, M. M. Mansour, G. C. Lie and E.
Cementi, J. Stat. Phys. \textbf{47} 209 (1987)

\bibitem {GKR07}C. Giardin{\`{a}}, J. Kurchan and F. Redig, J. Math. Phys.
\textbf{48} 033301 (2007)

\bibitem{SchmittmannZia} B. Schmittmann and R. K. P. Zia., ``Phase Transitions and Critical Phenomena vol 17,'' ed C. Domb and J. Lebowitz, Academic Press, London (1995).

\bibitem {long_range_exp} B. M. Law, R. W. Gammon, and J. V. Sengers, Phys.
Rev. Lett. \textbf{60} 1554 (1988)

\bibitem {DQD_review}W. G. van der Wiel, et al., Rev. Mod. Phys.
\textbf{75} 1 (2002)

\bibitem {ciliberto}S. Ciliberto, et al., arXiv:1301.4311 (2013)

\bibitem {DQD}T. Fujisawa, et al., Science \textbf{312} 1634  (2006). K\"{u}ng, B., et
al., Physical Review X \textbf{2} 011001 (2012)

\bibitem {fluct_heat_exchange}C. Jarzynski and D.~K.~W\'{o}jcik, Phys. Rev.
Lett. \textbf{92}, 230602 (2004)

\bibitem {our_coupled_sys}G. Bunin and Y. Kafri, J. Phys. A: Math. Theor. \textbf{46} 095002  (2013)

\bibitem{Tikhonenkov} I. Tikhonenkov,  A. Vardi, J.R. Anglin, and D. Cohen, Phys. Rev. Lett. {\bf 110}, 050401 (2013).


\bibitem {ILW09}A. Imparato, V. Lecomte, and F. van Wijland, Phys. Rev.
\textbf{80}, 011131 (2009)

\bibitem {BGL05}L. Bertini, D. Gabrielli and J. L. Lebowitz, J. Stat. Phys.
\textbf{121} 843 (2005)

\bibitem {KMP}C. Kipnis, C. Marchioro and E. Presutti, J. Stat. Phys. \textbf{27} 65 (1982)

\bibitem {DEHP93}B. Derrida, M.R. Evans, V. Hakim and V. Pasquier, J. Phys. A
\textbf{26}, 1493-1517 (1993)

\bibitem {gardiner}C. W. Gardiner, Handbook of stochastic methods for physics,
chemistry, and the natural sciences, Springer (1994)




\bibitem {BertiniPRL}L. Bertini, A. De Sole, D. Gabrielli, G. Jona-Lasinio,
and C. Landim, Phys. Rev. Lett. \textbf{87} 040601 (2001)

\bibitem {TKL_long}J. Tailleur, J. Kurchan, and V. Lecomte, J. Phys. A: Math.
Theor. \textbf{41} 505001 (2008)

\bibitem {our_numerics}G. Bunin, Y. Kafri, and D. Podolsky, EPL \textbf{99} 20002 (2012)

\bibitem {ASEP_transition}L. Bertini, A. De Sole, D. Gabrielli, G.
Jona-Lasinio, and C. Landim, J. Stat. Mech.  L11001 (2010)

\bibitem {ours_short}G. Bunin, Y. Kafri and D. Podolsky, J. Stat. Mech. L10001 (2012)

\bibitem {lr-corr_ph-trans} T. Bodineau, B. Derrida, V. Lecomte and F. van Wijland, J. Stat. Phys  \textbf{133} 1013 (2008)

\bibitem{BSGJL07}  L. Bertini, A. De Sole, D. Gabrielli, G. Jona-Lasinio, C. Landim, \texttt{arXiv:0705.2996} (2007)

\bibitem{ED04} C. Enaud and B. Derrida, J. Stat. Phys \textbf{114} 537 (2004)





\end{thebibliography}
\end{document}